\documentclass[manuscript]{aastex} 
\newcommand{\etal}{{\it et al.\/}}
\newcommand{\kmsM} {km~s$^{-1}$~Mpc$^{-1}$}

\begin{document}
\pagenumbering{arabic}

\title{Caltech Faint Galaxy Redshift Survey XV:\\
Classifications of Galaxies with $0.2 < z < 1.1$ in the Hubble Deep Field
(North) and its Flanking 
Fields \footnote{Based in part on observations obtained at the 	W.M. Keck
Observatory, which is operated jointly by the California Institute of
Technology and the University of California}}

\author{Sidney van den Bergh}
\affil{Dominion Astrophysical Observatory, Herzberg Institute of
Astrophysics, National Research Council of Canada, 5071 West Saanich Road,
Victoria, BC, Canada  V9E 2E7}
\email{sidney.vandenbergh@nrc.ca}

\author{Judith G. Cohen}
\affil{Palomar Observatory, Mail Stop 105-24, California Institute of
Technology, Pasadena, CA 91125}
\email{jlc@astro.caltech.edu}

\and

\author{Christopher Crabbe}
\affil{California Institute of Technology, Mail Stop 185--54, Pasadena, CA
91125}
\email{crabbe@its.caltech.edu}

\begin{abstract}

To circumvent the spatial effects of resolution on galaxy classification,
the images of 233 objects of known redshift in the Hubble Deep Field (HDF)
and its Flanking Fields (FF) that have redshifts in the range $0.20 < z <
1.10$ were degraded to the resolution that they would have had if they were
all located at a redshift $z= 1.00$. As in paper XIV of the present series,
the effects of shifts in rest wavelength were mitigated by using $R$-band
images for the classification of galaxies with $0.2 < z < 0.6$ and $I$-band
images for objects with redshifts $0.6 < z < 1.1$. A special effort was made
to search for bars in distant galaxies. The present data strongly confirm
the previous conclusion that the Hubble tuning fork diagram only provides a
satisfactory framework for the classification of galaxies with $z ~< 0.3$.
More distant disk galaxies are often difficult to shoehorn into the Hubble
classification scheme. The paucity of barred spirals and of grand-design
spirals at large redshifts is confirmed. It is concluded that the morphology
of disk galaxies observed at look-back times smaller than 3--4 Gyr differs
systematically from that of more distant galaxies viewed at look-back times
of 4--8 Gyr. The disks of late-type spirals at $z >0.5$ are seen to be more
chaotic than those of their nearer counterparts. Furthermore the spiral
structure in distant early-type spirals appears to be less well-developed
than it is in nearby early-galaxies.

\end{abstract}

% 235 becomes 233 when H369493_1317 is moved to z=1.238 and when
% one case of duplication is eliminated.

\section{Introduction}

The Hubble Space Telescope has, for the first time, allowed us to undertake
systematic imaging surveys \citep{williamsetal1996} of galaxies at large
redshifts. Furthermore, spectra obtained with the W. M. Keck 10-m telescope
\citep{cohenetal2000} have made it possible to determine redshifts (and hence
look-back times) for significant numbers of such distant galaxies. In recent
papers \citep{vdbetal1996,vdbetal2000,binchmannetal1998}
\footnote{\citet{binchmannetal1998} classify galaxies with respect to nine
fundamental type standards. One of us (SvdB) would have classified their Sbc
standard as Sb, their Scd standard as Sc I and their Ir standard (which
appears to have a central nucleus) as a spiral.} it was found that disk
galaxies at redshifts $\gtrsim$0.3 have morphologies that appear to differ
systematically from those of nearby galaxies. A question that presents
itself quite insistently is: Could the decrease in linear resolution with
increasing distance contribute significantly to these apparent systematic
changes of morphology with redshift? In an attempt to answer this question
we have degraded the images of all galaxies with redshifts 0.20 to 1.00 to
the appearance that they would have had at $z = 1.00$. (The images of 20
% 22 becomes 20 when problem cases are resolved
galaxies with $1.0 < z < 1.1$ were left unaltered.) As has already been
discussed in \citet{vdbetal2000} (henceforth vdB2000), 
all galaxies were classified at similar
rest wavelengths by comparing $R$-band images of galaxies having $0.20 < z <
0.60$ with $I$-band images of objects having $0.60 < z < 1.10$. For
statistical purposes the present data may be compared with the $B$-band
images of galaxies with $z \sim 0.0$ that are seen on the Palomar Sky
Survey. The redshifts for individual HST + FF galaxies in the present paper
are from \citet{cohenetal2000} and \citet{cohen2001}. Since only a single
redshift was available for five merging/interacting galaxies the total
number of images of galaxies examined (233) is slightly larger than the
total number of redshifts (228).
% 230 becomes 228 when total goes down to 233 from 235

For more detailed references to previous work on the morphology and
classification of distant galaxies the reader is referred to the excellent
review by \citet{abraham1999}.

\section{Effects of Resolution and of Noise}

``Postage stamp'' images of every individual galaxy with $0.20 < z < 1.00$
were manipulated in both brightness and in angular scale to mimic the
appearance that they would have had at $z = 1.00$. The images of galaxies
with $1.00 < z < 1.10$ were left untouched. For $z < 0.6$ the HST F606W
images were used, while the F814W images were employed for those galaxies in
the sample with $z > 0.6$. For many of the galaxies in the Flanking Fields,
no F606W images were available. These objects were therefore omitted from
the present study.

The angular scale for each galaxy with $z < 1.00$ was compressed by a factor
$f$ = (observed angular diameter) / (angular diameter at $z = 1.00$), using
a cosmology with H$_0$ = 60 \kmsM\ and  $\Omega_M = 0.3$. We then corrected
the intensity in the F606W images to take into account the wavelength
dependence of the difference in photon detection efficiency between the
WFPC2 instrument behind the F606W and F814 filters. Finally each image with
$z < 1.00$ was dimmed by a factor $S = [D_L (z=1)/ D_L(z)]^2$. Furthermore
we took into account the dependence on rest wavelength of the spectral
energy distribution of a galaxy; such objects are typically much brighter at
7500 \AA\ in the rest frame than they are at 4050 \AA. 

The procedure outlined above also suppresses the noise in the subimage
arising from instrumental contributions and from the sky by the same factor
$S$, which for low redshifts may be quite significant. At this point the
contrast between the galaxy and the background is artificially enhanced. To
restore the proper noise level, a set of images consisting of random
realizations of the noise characteristics of the F606W images in the three
WF CCDs and in the PC were generated. These were derived from the noise
properties measured in the original HST images in regions apparently free of
galaxies. Then, for galaxies with $z < 0.6$, these were added appropriately
to the compressed and rescaled images.\footnote{In principle this should
have been done for the ``postage stamps''  with $0.6 < z < 1.0$ as well, but
tests on the lower redshift subimages showed that restoring the correct
noise level had no significant effect for our purposes.} This then creates a
set of ``postage stamp'' images in which each galaxy appears as it would if
that object were situated at $z = 1.0$ in the image in which it was
originally observed.

Comparison of the new images with those used in vdB2000 showed
no cases in which our original classifications had to be revised. The only
(quite subtle) difference was that faint, possibly tidal, features in the
outer parts of some images were more difficult to see than they had been in
the original images. No significant differences were seen between the
morphologies of the main bodies of our program galaxies before, and after,
an appropriate amount of noise had been added. It is therefore concluded
that the classifications of vdB2000 are robust with regard to reasonable changes in the
noise level.

\section{Classification of the Images}

Table 1 lists the galaxies for which images degraded
to $z = 1.00$ were available from the survey of \citet{cohenetal2000}
augmented in \citet{cohen2001}. In an attempt to search for weak bars these
images were scrutinized over a larger dynamical range 
than had previously been explored for the classifications 
given in vdB2000. Classifications
for all objects were made by SvdB and are given on the 
DDO system \citep{vdb1960a,vdb1960b,vdb1960c}. This system is 
close to that adopted by \citet{sandagetammann1981}. The 
main difficulty encountered during the classification 
process was that a significant fraction of the galaxies  
at large redshifts look ``odd'' and were therefore often 
difficult (or impossible) to shoehorn into the 
classification bins of the Hubble system. For such objects a 
``p'' (peculiar) was added to their classifications. Since spiral 
structure in distant galaxies, which are viewed at large
look-back times, is less developed than it is in nearby 
galaxies, the morphology of spiral arms often can not be 
used as a classification parameter. Central concentration 
of light was therefore often used as the primary 
classification criterion for determining where a spiral
galaxy is located on the Sa -- Sb -- Sc sequence. Inspection of
the present images that were degraded to their appearance 
at $z = 1.00$ strengthens and confirms the previous conclusion 
\citep{vdbetal1996,vdbetal2000} that the morphology of 
late-type disk galaxies at large look-back times is more 
chaotic than that of nearer spirals of types  Sbc -- Sc. In
distant early-type disk galaxies the spiral arms are seen
to be less developed than they are in nearby Sa -- Sb galaxies.
As the images used to derive this conclusion had all been 
degraded to $z = 1.0$, the observed decrease in the 
strength of spiral structure with increasing $z$ cannot be
a resolution effect.

The surface density of galaxy images at the depth of the 
HST images of the HDF and FF
is quite high.  As a result, these fields contain many
apparent close pairs of galaxies, several of which were
shown in our previous paper to have widely discrepant redshifts, and hence
are
established as chance projections.
Because of this high probability of finding 
chance pairs we have, whenever practicable, 
used tidal deformation of images as a criterion for 
assigning objects to the ``merger'' class. Because we 
have been more conservative in designating objects to 
this class the fraction of ``mergers'' is smaller than it
was in vdB2000.
 
Our work provided many examples of late-type galaxies 
at high redshifts that appeared to be in a very early 
evolutionary state. With growing experience we have 
therefore become more confident in assigning objects 
to the ``proto-Sc'' class. This classification type is 
therefore now more common than it was in our previous 
paper. As a result some objects previously dubbed ``Sc 
pec'' have now been reclassified as ``proto-Sc''.
  
The fact that galaxies at large redshifts appear so
odd, and are difficult to fit into the Hubble scheme, 
adds significantly to the uncertainty of assigned Hubble
types. It is therefore a source of satisfaction to find
such good agreement between our original 
classifications of individual galaxies in vdB2000 and the present
classifications of images of these same objects degraded 
to $z = 1.0$. A blind comparison between the present new 
classifications of 99 objects that could be placed on the 
Hubble sequence E--Sa--Sb--Sc--Ir, and those which were previously
published for these same objects in vdB2000
% \citet{vdbetal2000} 
shows exact agreement for 49 objects, a difference
of 0.5 Hubble class for 23 objects, a difference of 1.0
Hubble class for 23 objects and a difference $>1.0$ Hubble
class for 4 objects. For three of these four discrepant
classifications the cause appears to have been that a small
fraction of the images classified in vdB2000
were not inspected over a sufficiently large dynamic
range. 

For this sample in common of 99 galaxies that could be 
placed on the Hubble sequence E--Sa--Sb--Sc--Ir there is thus
no evidence for any systematic difference in Hubble type
between the old classifications and the new classifications 
of images degraded to $z = 1.0$. Furthermore, there is no
evidence for any systematic dependence of the differences 
between the old and new classifications on redshift. Among
56 galaxies with $0.20 < z < 0.80$ the mean difference,
% 55 became 56 after check of problem cases
in the sense old type minus new type, is found to be $-0.04 \pm0.11$
Hubble classes. For the 41 galaxies on the Hubble 
sequence with $0.80 < z < 1.10$,
% 43 becomes 41 after check of problem cases 
the mean difference, in the 
sense old type minus new type, is $-0.08 \pm 0.12$ Hubble 
classes. Furthermore, the differences between old and new 
classification types does not appear to depend in a 
systematic way on Hubble type. On the basis of these results 
it is concluded that there are no significant differences
between Hubble types assigned in vdB2000
and those of degraded images in the present 
study. A tabulation of the frequencies of various classification
types of galaxies in Table 1 as a function of redshift is
shown in Table 2.

\section{Galaxy Statistics}

\subsection{Frequency of Barred Galaxies}

It was first noted that barred spirals appear to be deficient 
among galaxies at high redshift by \citet{vdbetal1996}. 
Subsequently this shortage of barred objects at
$z >0.5$ was confirmed (in both the northern and in the 
southern HDF) by \citet{abrahametal1999}. \citet{bothun2000}
suspected that the absence of SB galaxies at high 
redshifts might have been due to bandshift effects that
result from the fact that bars appear to be more
frequent in the $I$-band \citep{eskridgeetal2000} than they 
are in the $B$-band. However, the observations of bar 
frequencies at various redshifts by vdB2000, 
which were obtained at almost constant rest  
wavelength, appear to rule out this explanation.

In the present investigation special attention was 
paid to the possible presence of bars by inspecting 
each program galaxy over the widest possible dynamic 
range. Out of 233 images, only a single one was found 
to be a pure barred spiral of type SB, while one 
was an intermediate type object classified as S(B).
Furthermore three galaxies were classified S(B?). The
corresponding percentages are 0.4\% SB, 0.4\% S(B) and
1.3\% S(B?), respectively. For comparison, \citet{sandagetammann1981} 
found that 22\% of all Shapley-Ames galaxies, 
which have $z \sim0.0$, are barred. These data suggest that 
the fraction of all galaxies that are barred is at least 
an order of magnitude lower in the distant HDF + FF sample 
than it is in the nearby Shapley-Ames sample. It is noted 
in passing that the only two certain barred galaxies
in the present sample [which were classified SB and S(B)]
have $z < 0.5$. Nevertheless, it is puzzling that the fraction 
of barred spirals at all redshifts appears to be so much lower 
on HST images than it is on photographs of nearby galaxies. Simulations to 
investigate this problem are currently being undertaken by
\citet{abrahamvdbprep}.

Since the Universe is expanding, galaxies were once
closer together than they are at the present time. As
a result, tidal interactions and mergers are expected
to be more frequent at high redshifts than they are 
at the present time. Evidence based on galaxy morphology
in favor of this view was first provided by observations of galaxies in the 
Hubble Deep Field \citep{vdbetal1996}. Possibly 
the high frequency of tidal interactions in the early 
universe also contributed to the shortage of large 
well-developed disks in galaxies at $z >> 1$, which was 
already noted above. Using the present (rather strict) 
definition of mergers, the data in Table 2 show that 
11 out of 116 (~9\%) of all galaxies with $0.70 < z < 1.00$ 
are classified as either ``Merger'' or ``Peculiar/merger''. 
From inspection of the entire data set, one of us (SvdB) 
has the impression that the rate of multiple mergers 
may be increasing faster with redshift that the rate of 
binary mergers.

\subsection{Changes in the Relative Frequencies of Galaxy Types
     as a Function of Redshift}

Table 2 shows the frequency distributions of different
morphological types in the present sample as a function of
redshift. The data in this table may be compared to those
of the northern Shapley-Ames galaxies \citep{vdb1960c}
given in vdB2000. The Shapley-Ames galaxies, which are located 
at $z \sim0.0$, were classified 
on reproductions of the $B$ images of the Palomar Sky Survey. 
They have therefore been observed at approximately 
the same rest wavelength as the HST images of galaxies 
with $z > 0.20$. Since galaxies of types E, Sa and Sb 
have rather similar luminosity functions their relative 
frequencies (as a function of redshift) should not be 
strongly affected by redshift-dependent selection effects. However, the
present
luminosities of late-type spirals are, on average, 
significantly lower than those of galaxies having earlier 
Hubble types [see Fig. 1 of \citet{vdb1998}]. In the
absence of evolutionary effects, high-redshift samples of 
galaxies might therefore be biased against objects of 
types Sc and Ir. It should, however, be emphasized that this 
bias against late-type galaxies will be lessened (or might
even be reversed) if such objects exhibit significant 
luminosity evolution. \citet{lillyetal1995} and \citet{cohen2000}
have, for example, demonstrated that galaxies with $0.5 < z < 0.75$ might
have brightened by as much as ~1 mag. On the other hand,  
\citet{carollolilly2001} have more recently noted that the 
high metallicity of late-type galaxies with $0.5 < z < 1.0$ 
makes it improbable that these objects are luminosity-enhanced dwarfs. 

The data in Table 2 suggest that the fraction of
late-type (Sc + Sc/Ir + Ir) galaxies decreases with
increasing redshift. For the redshift bins $z = 0.20-0.49$,
$0.50-0.79$ and $0.80-1.09$ the percentage of late-type galaxies
is found to be 19\%, 12\% and 8\%, respectively. Taken
at face value, this trend suggests that the fraction of
late-type galaxies decreases with increasing redshift.
Such an effect could, if real, be explained by assuming 
that a large fraction of all Sc galaxies have not yet
been fully assembled, or are still classified as ``proto-Sc'', 
at $z >~0.5$.  Alternatively, if
luminosity evolution is unimportant, this effect might 
be due to a distance-dependent selection effect which 
results from the fact that late-type galaxies are (on
average) less luminous than early-type galaxies [see
Fig. 1 of \citet{vdb1998}].

The present sample is so small that bins containing
objects in a small range of redshifts and morphological 
types provide pitifully small statistical samples of 
galaxies. Furthermore the distribution of morphologies at 
a given redshift might be affected by density enhancements
as the line of sight passes through sheets or clusters of
galaxies. Intersection of the line of sight towards the HDF
with a populous group of galaxies at $z \sim0.68$ is, 
for example, responsible for 
an excess of E galaxies in the $z = 0.60 - 0.69$ redshift 
bin. As a result of the vagaries of small-number statistics, 
and the effects of local density enhancements on galaxy 
morphology, it would be unwise to draw very strong 
conclusions from the present data about the morphological 
evolution of galaxies over time.

The apparent increase in the number of ellipticals with $z > 0.9$ might 
be spurious, even if it is not due to the vagaries of small number 
statistics. This is so because any small compact high 
surface brightness object with regular oval
isophotes will be classified as an ``elliptical''. This includes
the class of galaxies denoted as ``blue compact galaxies''
[see, for example, \citet{phillipsetal1997}, \citet{guzmanetal1997}],
a local example of which might be NGC 6789.
These two groups are not easily distinguished morphologically
in this redshift regime. However, from the spectra and SEDs
of the set of galaxies under consideration here
given in earlier papers in this series, we are confident that many of these
galaxies are not classical ellipticals, but are instead strongly
star forming blue galaxies which are compact.

\section{Morphological Evolution of Galaxies}

Since the majority of elliptical galaxies are 
believed to have formed at $z >1.0$, the present galaxy
sample is not expected to contain many proto-ellipticals. 
Nevertheless, some compact pairs and 
groupings of early-type galaxies might, in the 
fullness of time, merge into run-of-the-mill 
ellipticals. A few E and Sa galaxies in the present
sample exhibit faint chaotic outer structures that 
might have resulted from such earlier mergers.

Many of the early-type (Sa, Sab) spirals in 
our sample exhibit morphological peculiarities.
The most common of these are: (1) underdeveloped 
spiral structure, (2) warped disks and (3) off-center nuclear bulges.
A large fraction of the late-type (Sbc, Sc, 
Sc/Ir) galaxies in our sample of high redshift 
galaxies exhibit morphological peculiarities that
distinguish them from their nearby counterparts. The
disks of distant Sc galaxies tend to be much more
chaotic than those of nearby luminous galaxies which
generally exhibit a well-ordered spiral structure.
This effect is particularly significant because the
sample of distant late-type galaxies is biased in
favor of objects of above-average luminosity. Such
galaxies are expected (van den Bergh 1960a) to 
exhibit particularly long and regular spiral arms.
Distant single Sc spirals sometimes  have one strongly
dominant spiral arm. Among nearby galaxies such single-arm objects are
usually objects that have been tidally 
perturbed. Furthermore, the disks of proto-Sc galaxies, 
in which spiral structure is not yet well developed, 
are often seen to contain luminous blue knots (super-associations). 
In other words, huge localized bursts of 
star formation appear to occur before a well-ordered
spiral pattern develops. If the regions in which such
super-associations occur have significant over densities
they could experience strong dynamical friction and
spiral towards the galactic center \citep{noguchi1998},
where they might contribute to the formation of a 
galactic bulge.

Figure 1 shows a good example of an Sc I galaxy at 
$z = 1.01$ that is just beginning to form well-ordered 
spiral structure. Note that the arms of this object are 
much patchier and more chaotic than those of typical
nearby Sc I galaxies. The luminous patches (super-
associations) in this object are seen to be much more
luminous than those that occur in typical Sc I galaxies 
at $z \sim 0.0$. A color image of an even more primitive
late-type galaxy is shown as Plate 9 of \citet{vdbetal1996}. 
This object exhibits what appears to be a disk, 
in which ~10 bright blue knots (super-associations) are
embedded. A single slightly off-center red knot may be
the (slightly older) nuclear bulge of this probable
protogalaxy. Figure 2 shows an example of an object
at z = 1.06 that might be a proto-Sc with a single arm 
that contains a super-association. Alternatively this
object might be interpreted as the final phase of a
merger. Spectroscopic observations will be required to
distinguish between these two alternatives. Finally,
Figure 3 shows what may be another example of a 
proto-galactic disk in which a nucleus, and four other
bright knots, appear to be embedded.

\section{Conclusions}

The present data have allowed us to study the evolution
of galaxy morphology over the range $0.2 < z < 1.1$ at an
effective wavelength that is insensitive to redshift, and
at a spatial resolution that is independent of $z$. The main 
conclusion that can be drawn from this work is that the 
morphology of disk galaxies has evolved rapidly over the
last 8 Gyr. At look-back times $>4$ Gyr, late-type spirals
generally have a much more chaotic structure than do
nearby galaxies of type Sbc and Sc. Furthermore, the arms
of such distant spirals tend to be patchy and are also
often asymmetric. At look-back times $>4$ Gyr, early-type
spiral galaxies appear to have less well-developed spiral 
structure than do their counterparts that are viewed at
look-back times of less than 3--4 Gyr. Apparently, insufficient 
time was available for spiral structure to develop in such early-type 
galaxies. Among distant galaxies in the Hubble Deep Field the intrinsic
frequency 
of barred spirals appears to be at least an order of
magnitude lower than it is for nearby galaxies at $z \sim0$. 
It is concluded that the \citet{hubble1936} tuning fork 
scheme is only appropriate for the classification of
galaxies at $z ~<0.3$ which are viewed at look-back times
of less than 3--4 Gyr.

\acknowledgements

We are indebted to Roger Blandford and to David Hogg for their assistance
and advice during the course of the present investigation. We also thank Bob
Abraham for his help with the figures.

\clearpage

\begin{deluxetable}{lclll}
%\tablewidth{307pt}
\tablecaption{\sc{Classifications of Galaxies} }
\tablehead{\colhead{ID\tablenotemark{a}} & \colhead{Redshift\tablenotemark{b}} 
& \colhead{$R$\tablenotemark{c}} 
& \colhead{Classification} & \colhead{Comments} \\
\colhead{ } & \colhead{ } & \colhead{(Mag)}  \\
}
\startdata
F36438\_1357 &   0.201 & 20.65 & Sb: & On edge of image \\
% $ 0.201 $ & Sb:          & $36$ m $43$ s $75$ & $13$'$56$"$7$ &
% \tablenotemark{a}  \\
F36510\_0938  &  0.205  &  21.27 & Ir & Edge-on \\
% $ 0.205 $ & Ir           & $36$ m $51$ s $07$ & $09$'$38$"$6$ &
% \tablenotemark{e}  \\
F36545\_1014 &   0.224   & 23.86 & Sa & \\
% $ 0.224 $ & Sa           & $36$ m $54$ s $57$ & $10$'$14$"$6$ & \\
H36560\_1329 &  0.271 &  23.80 &  Ir \\
% $ 0.271 $ & Ir           & $36$ m $56$ s $10$ & $13$'$29$"$6$ &
% \tablenotemark{i}  \\
%   There is no text for tablenote i.
H36580\_1300 &  0.320 & 22.04 &  Sabp(t?)  &  Double nucleus \\
% $ 0.320 $ & Sabp(t?)     & $36$ m $58$ s $04$ & $13$'$00$"$4$ & \\
F36563\_1209 & 0.321 &  23.22 &  Sb - Ir  &  Edge-on  \\
% $ 0.321 $ & Sb - Ir      & $36$ m $56$ s $39$ & $12$'$09$"$3$ &
% \tablenotemark{e}  \\
H36470\_1236 &  0.321 &  20.62 &  Sb:p &  \\
% $ 0.321 $ & Sb:p         & $36$ m $47$ s $00$ & $12$'$36$"$9$ & \\
H36587\_1252 &  0.321 & 20.99 &  SBcp &  One long arm \\
% $ 0.321 $ & SBcp         & $36$ m $58$ s $73$ & $12$'$52$"$4$ &
% \tablenotemark{q}  \\
H36508\_1255 &  0.321 & 22.27 &   Sp &  Edge-on \\
% $ 0.321 $ & Sp           & $36$ m $50$ s $82$ & $12$'$55$"$8$ &
% \tablenotemark{e}  \\
H36551\_1311 &  0.321 & 23.58 &  Merger? \\
% $ 0.321 $ & Merger?      & $36$ m $55$ s $30$ & $13$'$11$"$3$ & \\
F36458\_1325 &  0.321  & 20.71  & Scp   &  Rudimentary spiral structure \\
% $ 0.321 $ & Scp          & $36$ m $45$ s $86$ & $13$'$25$"$7$ &
% \tablenotemark{h}  \\
H36526\_1219 & 0.401 & 23.11 &  E4p & Embedded in asymmetric nebulosity \\
% $ 0.401 $ & E4p          & $36$ m $52$ s $66$ & $12$'$19$"$7$ &
% \tablenotemark{l}  \\
H36516\_1220 &   0.401   & 21.45 & Sabp &  Binary nucleus \\
% $ 0.401 $ & Sabp         & $36$ m $51$ s $69$ & $12$'$20$"$2$ &
% \tablenotemark{k}  \\
F36410\_0949  &  0.410 & 20.38 & Ir/Merger? & \\
% $ 0.412 $ & Ir/Merger?   & $36$ m $41$ s $30$ & $09$'$48$"$8$ & \\
H36472\_1230 &  0.421 & 22.63 &  S(B?)bp & Edge-on \\
% $ 0.421 $ & S(B?)bp      & $36$ m $47$ s $28$ & $12$'$30$"$7$ &
% \tablenotemark{e}  \\
H36419\_1205 & 0.432  & 20.82 & Scp &   \\
% $ 0.432 $ & Scp          & $36$ m $41$ s $95$ & $12$'$05$"$4$ & \\
H36513\_1420 &  0.439  & 23.22 &  Sb(?)p   &  \\
% $ 0.439 $ & Sb(?)p       & $36$ m $51$ s $37$ & $14$'$20$"$9$ & \\
F36454\_1325 &  0.441 & 22.33 & Sap & Asymmetric \\
% $ 0.441 $ & Sap          & $36$ m $45$ s $80$ & $13$'$25$"$8$ &
% \tablenotemark{f}  \\
H36465\_1203 &  0.454 & 24.32 &  ?   &  Tidal debris? \\
% $ 0.454 $ & ?            & $36$ m $46$ s $51$ & $12$'$03$"$5$ &
% \tablenotemark{ad} \\
H36429\_1216 &  0.454 &  20.51 & Sc I:  &   \\
% $ 0.454 $ &         & $36$ m $42$ s $93$ & $12$'$16$"$4$ & \\
H36448\_1200 &  0.457 & 22.85 &  Sbcp & Rudimentary spiral structure \\
% $ 0.457 $ & Sbcp         & $36$ m $44$ s $83$ & $12$'$00$"$1$ &
%\tablenotemark{h}  \\
H36519\_1209 &  0.458 & 22.75 & Sp(t?) \\
% $ 0.458 $ & Sp(t?)       & $36$ m $51$ s $99$ & $12$'$09$"$6$ & \\
H36594\_1221 &  0.472 &  23.53 & Sa + Sp & Spectrum probably refers to \\
       &   &   &   &  combined light \\
% $ 0.472 $ & Sa + Sp      & $36$ m $59$ s $41$ & $12$'$21$"$5$ &
% \tablenotemark{ae} \\
H36501\_1239 & 0.474 & 20.43 & Sc?p &  Binary nucleus \\
% $ 0.474 $ & Sc?p         & $36$ m $50$ s $19$ & $12$'$39$"$8$ &
% \tablenotemark{k}  \\
H36569\_1302 &   0.474 & 23.69 & E0/star & \\
% $ 0.474 $ & E0/star      & $36$ m $56$ s $89$ & $13$'$01$"$5$ & \\
H36496\_1257 &  0.475  & 21.91 &  Pec & Compact and asymmetric \\
% $ 0.475 $ & Pec          & $36$ m $49$ s $63$ & $12$'$57$"$6$ &
% \tablenotemark{j}  \\
H36572\_1259 &  0.475 & 21.07 &  S(B)cp  & Perhaps proto?-SBc \\
% $ 0.475 $ & S(B)cp       & $36$ m $57$ s $27$ & $12$'$59$"$5$ &
% \tablenotemark{o}  \\
H36497\_1313 & 0.475 & 21.46 &  Sbp    &   Asymmetric \\
% $ 0.475 $ & Sbp          & $36$ m $49$ s $70$ & $13$'$13$"$0$ &
% \tablenotemark{f}  \\
H36480\_1309  &  0.476   & 20.43 & E0p & Asymmetric envelope \\
% $ 0.476 $ & E0p          & $36$ m $48$ s $07$ & $13$'$09$"$0$ &
% \tablenotemark{d}  \\
H36493\_1311 &  0.477  & 21.97 &   E2 \\
% $ 0.477 $ & E2           & $36$ m $49$ s $36$ & $13$'$11$"$2$ & \\
H36415\_1200 &   0.483   & 25.03 & Pec & \\
% $ 0.483 $ & Pec          & $36$ m $41$ s $62$ & $12$'$00$"$5$ & \\
H36508\_1251 & 0.485 & 23.15 & proto-Sc  &  \\
% $ 0.485 $ & proto-Sc     & $36$ m $50$ s $83$ & $12$'$51$"$5$ & \\
F36446\_1304 & 0.485 &  21.14 & Sbp &  Proto-Sb? \nl
% $ 0.485 $ & Sbp          & $36$ m $44$ s $61$ & $13$'$04$"$6$ &
% \tablenotemark{s}  \\
F36427\_1306 & 0.485 & 22.02 &  Sab & Edge-on  \nl
% $ 0.485 $ & Sab          & $36$ m $42$ s $71$ & $13$'$06$"$7$ &
% \tablenotemark{e}  \\
H36528\_1404 &  0.498  & 23.45 &  Pec/Merger \\
% $ 0.498 $ & Pec/Merger   & $36$ m $52$ s $84$ & $14$'$04$"$8$ & \\
H36465\_1151 &  0.503  & 22.00 &  E0 &  \nl
% $ 0.503 $ & E0           & $36$ m $46$ s $51$ & $11$'$51$"$3$ & \\
F36397\_1009  &  0.509  & 21.59 & Sbp & Asymetric \\
% $ 0.509 $ & Sbp          & $36$ m $39$ s $68$ & $10$'$09$"$7$ &
% \tablenotemark{f}  \\
H36549\_1314 &  0.511 & 23.81 & ?  &  Edge-on \\ 
% $ 0.511 $ & ?            & $36$ m $54$ s $96$ & $13$'$14$"$8$ &
% \tablenotemark{e}  \\
H36489\_1245 &  0.512 &  23.48 & Ir: &  \\
% $ 0.512 $ & Ir:          & $36$ m $48$ s $98$ & $12$'$45$"$8$ & \\
H36536\_1417 & 0.517 & 23.36 & Sab/S0  &  Edge-on \\ 
% $ 0.517 $ & Sab/S0       & $36$ m $53$ s $65$ & $14$'$17$"$6$ &
% \tablenotemark{e}  \\
F36516\_1052  &  0.518  & 23.04 & Ir & \\
% $ 0.518 $ & Ir           & $36$ m $51$ s $64$ & $10$'$52$"$3$ & \\
H36566\_1245 &  0.518  & 20.06 &  Sbp & Smooth disk with little star formation \\
% $ 0.518 $ & Sbp          & $36$ m $56$ s $62$ & $12$'$45$"$5$ &
% \tablenotemark{m}  \\
F36213\_1417 &   0.520 & 20.73 & Sa & \\   
% $ 1.014 $ & Sa           & $36$ m $21$ s $33$ & $14$'$17$"$1$ & \\
% typo in redshift, next previous object is at 1.014
H36569\_1258 &  0.520 &  23.84 &  E5/Sa &  \nl
%$ 0.520 $ & E5/Sa        & $36$ m $56$ s $90$ & $12$'$58$"$0$ & \\
F36425\_1518 &   0.533 & 21.79 & Sbc &  \\
% $ 0.845 $ & Sbc          & $36$ m $42$ s $55$ & $15$'$18$"$4$ & \\
% typo in redshift
H36414\_1142 & 0.548 & 23.51 &  Merger  &  Merger of Sap + Sap + ? + Sab: \\
% $ 0.548 $ & Merger       & $36$ m $41$ s $43$ & $11$'$42$"$5$ &
% \tablenotemark{ac} \\
F36429\_1030 &   0.551 & 21.82 &  E+E Merger?  & \\
% $ 0.551 $ & E+E Merger?  & $36$ m $43$ s $01$ & $10$'$30$"$2$ & \\
H36442\_1247 &  0.555 &  21.40 &  Sc:p   &   \\
% $ 0.555 $ & Sc:p         & $36$ m $44$ s $20$ & $12$'$47$"$8$ & \\
H36439\_1250 &  0.557 & 20.84 &  Pec/Merger?  &    \\
% $ 0.557 $ & Pec/Merger?  & $36$ m $43$ s $97$ & $12$'$50$"$1$ & \\
H36517\_1353 &  0.557 & 21.08 &  Sp &   \\
% $ 0.557 $ & Sp           & $36$ m $51$ s $77$ & $13$'$53$"$7$ & \\
H36452\_1142 &  0.558 & 24.00 & ``Tadpole'' & \\
% $ 0.558 $ & ``Tadpole''  & $36$ m $45$ s $30$ & $11$'$42$"$9$ & \\
H36555\_1359 &  0.559 &  23.74 &  Sbc(p?) & Edge-on, Asymmetric? \\
% $ 0.559 $ & Sbc(p?)      & $36$ m $55$ s $55$ & $13$'$59$"$8$ &
% \tablenotemark{e} \tablenotemark{f?} \\
H36519\_1400 &  0.559 & 23.03 & Sbcp & Edge-on, Asymmetric \\
% $ 0.559 $ & Sbcp         & $36$ m $51$ s $96$ & $14$'$00$"$7$ &
% \tablenotemark{e} \tablenotemark{f}   \\
H36534\_1234 &  0.560  & 22.78 & Sb/Ir &  Multiple nuclei \\
% $ 0.560 $ & Sp/Ir        & $36$ m $53$ s $43$ & $12$'$34$"$3$ & \\
H36571\_1225 &  0.561 & 22.36 &  Scp  & One-armed spiral \\ 
% $ 0.561 $ & Scp          & $36$ m $57$ s $18$ & $12$'$25$"$9$ &
% \tablenotemark{n}  \\
H37005\_1234 &  0.563 & 21.43 & E0 \\
% $ 0.563 $ & E0           & $37$ m $00$ s $53$ & $12$'$34$"$7$ & \\
H36554\_1402 &  0.564 & 23.08 &  Sc/Ir  &  Edge-on \\
% $ 0.564 $ & Sc/Ir        & $36$ m $55$ s $49$ & $14$'$02$"$6$ &
% \tablenotemark{e}  \\
H36413\_1141 & 0.585 & 21.91 &  Sa   & \\
% $ 0.585 $ & Sa           & $36$ m $41$ s $34$ & $11$'$40$"$8$ & \\
H36389\_1219 & 0.609  &  22.14 &    Pec & \\
% $ 0.609 $ & Pec          & $36$ m $38$ s $99$ & $12$'$19$"$7$ & \\
H36471\_1414 &   0.609 & 23.92 & E5 + E1 & Spectrum probably refers to combined light \\
% $ 0.609 $ & E5 + E1      & $36$ m $47$ s $16$ & $14$'$14$"$4$ &
% \tablenotemark{ae} \\
F36244\_1454 & 0.628  &  20.34 &     E2 \\
% $ 0.628 $ & E2           & $36$ m $24$ s $41$ & $14$'$54$"$1$ & \\
F36249\_1252 & 0.631 & 22.76 &  Sab/S0 & \\
% $ 0.631 $ & Sab/S0       & $36$ m $24$ s $95$ & $12$'$52$"$1$ & \\
F36384\_1312 & 0.635  &  22.27 &  Sabp & Asymmetric \\
% $ 0.635 $ & Sabp         & $36$ m $38$ s $48$ & $13$'$12$"$9$ &
% \tablenotemark{f}  \\
F37163\_1432 & 0.635    & 22.50 & Sbp & Asymmetric \\
% $ 0.635 $ & Sbp          & $37$ m $16$ s $34$ & $14$'$32$"$8$ &
% \tablenotemark{f}  \\
F36287\_1357 & 0.639  & 23.05  & Sa: & \\
% $ 0.639 $ & Sa:          & $36$ m $28$ s $73$ & $13$'$57$"$8$ & \\
F36247\_1510 & 0.641 & 20.41 &   Sab & Rudimentary spiral structure \\
% $ 0.641 $ & Sab          & $36$ m $24$ s $71$ & $15$'$10$"$4$ &
% \tablenotemark{h}  \\
H36538\_1254 & 0.642  & 20.95 & (proto?)Sc & \\
% $ 0.642 $ & (proto?)Sc   & $36$ m $53$ s $88$ & $12$'$54$"$0$ & \\
F36248\_1438 &   0.642  & 21.59 & E1 & \\
% $ 0.642 $ & E1           & $36$ m $24$ s $91$ & $14$'$38$"$7$ & \\
F36254\_1519 &  0.642  &  21.82 &   Sbp &  Asymmetric \\
% $ 0.642 $ & Sbp          & $36$ m $25$ s $45$ & $15$'$19$"$6$ & \\
F36481\_1102 & 0.650 &  22.58 &  Sb & \\
% $ 0.650 $ & Sb           & $36$ m $48$ s $12$ & $11$'$02$"$2$ & \\
F37080\_1246 & 0.654 & 21.80 &  Sap & Asymmetric \\
% $ 0.654 $ & Sap          & $37$ m $08$ s $08$ & $12$'$46$"$8$ &
% \tablenotemark{f}  \\
F36250\_1341 & 0.654 &  24.32 &  E0/Star & \\
% $ 0.654 $ & E0/Star      & $36$ m $25$ s $07$ & $13$'$41$"$8$ & \\
F37072\_1214 & 0.655  &  22.19 &  Sab & \\ 
% $ 0.655 $ & Sab          & $37$ m $07$ s $25$ & $12$'$14$"$1$ & \\
F37213\_1120 &  0.656   & 22.22 & S(B?)c & \\ 
% $ 0.656 $ & S(B?)c       & $37$ m $21$ s $40$ & $11$'$20$"$3$ & \\
F37060\_1340  & 0.672 & 23.25 & S: & \\
% $ 0.672 $ & S:           & $37$ m $06$ s $04$ & $13$'$40$"$4$ & \\
F37171\_1122 &   0.676  & 22.98 & E1/Star & \\
% $ 0.676 $ & E1/Star      & $37$ m $17$ s $15$ & $11$'$22$"$6$ & \\
H36471\_1213 &   0.677 & 24.63 &  Sap + Sa/E3 & Asymmetric envelope, \\
     &         &   &    & spectrum probably refers to combined light \\
% $ 0.677 $ & Sap + Sa/E3  & $36$ m $47$ s $10$ & $12$'$12$"$5$ &
% \tablenotemark{d} \tablenotemark{ae} \\
F36588\_1434 & 0.678 & 20.85 &    Sb: & \\
% $ 0.678 $ & Sb:          & $36$ m $58$ s $82$ & $14$'$34$"$8$ & \\
H36459\_1201 & 0.679   & 23.88 & S:p &  Edge-on \\
% $ 0.679 $ & S:p          & $36$ m $45$ s $96$ & $12$'$01$"$3$ &
% \tablenotemark{e}  \\
H36502\_1245 &  0.680  & 21.74 &  E3/S0 & \\
% $ 0.680 $ & E3/S0        & $36$ m $50$ s $26$ & $12$'$45$"$7$ & \\
F36362\_1319 & 0.680  & 22.20 &  Sb:p & \\
% $ 0.680 $ & Sb:p         & $36$ m $36$ s $28$ & $13$'$19$"$8$ & \\
F36580\_1137 & 0.681 & 23.00  &  E1 & \\
% $ 0.681 $ & E1           & $36$ m $58$ s $06$ & $11$'$37$"$8$ & \\
H36475\_1252 &   0.681 & 24.26 & Sbp & Asymetric \\
% $ 0.681 $ & Sbp          & $36$ m $47$ s $55$ & $12$'$52$"$7$ &
% \tablenotemark{f}  \\
F36481\_1002 & 0.682  & 21.92 &  Sbp & \\
% $ 0.682 $ & Sbp          & $36$ m $48$ s $18$ & $10$'$02$"$5$ & \\
H36586\_1221 & 0.682 &  23.40 &   E3p & Non-elliptical isophotes \\
% $ 0.682 $ & E3p          & $36$ m $58$ s $63$ & $12$'$21$"$8$ &
% \tablenotemark{p}  \\
F36278\_1449 &   0.680 & 21.64 & Sa & \\
% $ 0.682 $ & Sa           & $36$ m $27$ s $89$ & $14$'$49$"$0$ & \\
%
% F36243\_1455  &  0.628 &   20.34 & E2 & \\  
% $ 0.682 $ & E2           & $36$ m $24$ s $41$ & $14$'$54$"$1$ & \\
% typo in redshift, double entry, see z=0.628
%
F36243\_1525 & 0.682  & 22.78 &   E2/Sa & \\
% $ 0.682 $ & E2/Sa        & $36$ m $24$ s $31$ & $15$'$25$"$3$ & \\
F36454\_1523 &  0.683  &  22.06 &   E4 & \\
% $ 0.683 $ & E4           & $36$ m $45$ s $42$ & $15$'$23$"$0$ & \\
F36499\_1058 & 0.684 &  22.63 &    Sp & Arms under-developed \\
% $ 0.684 $ & Sp           & $36$ m $49$ s $93$ & $10$'$58$"$6$ &
% \tablenotemark{c}  \\
F37113\_1545 & 0.692  &  22.43 &  Ir & \\
% $ 0.692 $ & Ir           & $37$ m $11$ s $36$ & $15$'$45$"$1$ & \\
F37069\_1208 & 0.693  & 24.13 &   E0/Sa & \\
% $ 0.693 $ & E0/Sa        & $37$ m $06$ s $93$ & $12$'$08$"$0$ & \\
F36290\_1346 & 0.693 & 23.02 &   E2 & \\
% $ 0.693 $ & E2           & $36$ m $29$ s $01$ & $13$'$46$"$6$ & \\
F36427\_1503 & 0.698  &  23.17 & Ir/Pec & Edge-on \\
% $ 0.698 $ & Ir/Pec       & $36$ m $42$ s $73$ & $15$'$02$"$7$ &
% \tablenotemark{e}  \\
F36415\_0902 & 0.713   &  22.31 & Sc: & \\
% $ 0.713 $ & Sc:          & $36$ m $41$ s $53$ & $09$'$02$"$0$ & \\
F37017\_1143  &  0.744 & 22.10 & Sa & \\
% $ 0.744 $ & Sa           & $37$ m $01$ s $77$ & $11$'$44$"$0$ & \\
F37020\_1517 & 0.744 & 23.56 &   Sb: & \\
% & $ 0.744 $ & Sb:          & $37$ m $02$ s $07$ & $15$'$17$"$3$ & \\
F37036\_1353 &  0.745   & 21.63  & Sb: & \\
% $ 0.745 $ & Sb:          & $37$ m $03$ s $80$ & $13$'$53$"$1$ & \\
F37108\_1059 & 0.747   & 24.25 &  ? + ? & Combined spectrum of two LSB objects \\ 
% $ 0.747 $ & ? + ?        & $37$ m $10$ s $91$ & $10$'$59$"$0$ &
% \tablenotemark{ag} \\
F36297\_1329 &   0.748 & 23.03 & S:p & Edge-on \\
% $ 0.748 $ & S:p          & $36$ m $29$ s $70$ & $13$'$25$"$0$ &
% \tablenotemark{e}  \\
F36522\_0957  & 0.750 & 23.07 & Sabp & Asymetric \cr
% $ 0.752 $ & Sabp         & $36$ m $52$ s $24$ & $09$'$57$"$6$ &
% \tablenotemark{f}  \\
% assumed confusion in ID rather than typo in redshift, see F36296_1324
H36498\_1242 & 0.751 & 24.38 &  Ir & Edge-on \\
% $ 0.751 $ & Ir           & $36$ m $49$ s $86$ & $12$'$42$"$3$ &
% \tablenotemark{e}  \\
F36275\_1418 & 0.751  &  22.37 &    Sb & Edge-on, Spectrum might refer to Ir? companion \\
% $ 0.751 $ & Sb           & $36$ m $27$ s $55$ & $14$'$18$"$8$ &
% \tablenotemark{e} \tablenotemark{af} \\
H36436\_1218 & 0.752 &   22.56 &   Sa & \\
% $ 0.752 $ & Sa           & $36$ m $43$ s $61$ & $12$'$18$"$1$ & \\
F37074\_1356 &  0.752   & 23.65  &  E2/Sa & \\
% $ 0.752 $ & E/Sa         & $37$ m $07$ s $42$ & $13$'$56$"$7$ & \\
H36494\_1406 & 0.752  &  21.95 &   Sa & \\
% $ 0.752 $ & Sa           & $36$ m $49$ s $90$ & $14$'$06$"$6$ & \\
F37058\_1317 & 0.753  & 21.95 &   Sc:p & No spiral arms \\
% $ 0.753 $ & Sc:p         & $37$ m $05$ s $81$ & $13$'$17$"$2$ &
% \tablenotemark{u}  \\
H36487\_1318 & 0.753   &  22.87 & proto-Sc & \\ 
% $ 0.753 $ & proto-Sc     & $36$ m $48$ s $77$ & $13$'$18$"$4$ & \\
F37061\_1332 &   0.753 & 21.85 & Sbp &  No spiral arms \\
% $ 0.753 $ & Sbp          & $37$ m $06$ s $19$ & $13$'$32$"$7$ &
% \tablenotemark{u}  \\
F36297\_1324 & 0.758   & 23.25 & Sap & Binary nucleus ? \\
% $ 0.758 $ & Sap          & $36$ m $29$ s $70$ & $13$'$25$"$0$ &
% \tablenotemark{k?} \\
F36299\_1440  &  0.762  & 22.81 & Sa:p & \\
% $ 0.762 $ & Sa:p         & $36$ m $29$ s $91$ & $14$'$40$"$9$ & \\
F36598\_1449 & 0.762  &  21.62 & Merger & \\
% $ 0.762 $ & Merger       & $36$ m $59$ s $89$ & $14$'$49$"$8$ & \\
H36438\_1142 & 0.765  &   21.26 &  E/Sa & Asymmetric envelope \\ 
% $ 0.765 $ & E/Sa         & $36$ m $43$ s $81$ & $11$'$42$"$9$ &
% \tablenotemark{d}  \\
F36379\_0922 &  0.767 & 21.43 &  S(B?)b(p?) & Earliest phase of bar formation? \\
% $ 0.767 $ & S(B?)b(p?)   & $36$ m $37$ s $97$ & $09$'$22$"$0$ &
% \tablenotemark{aa} \\
F37115\_1042 & 0.778  & 21.97 &   Sbcp & \\
% $ 0.778 $ & Sbcp         & $37$ m $11$ s $59$ & $10$'$42$"$4$ & \\
F37015\_1129 & 0.779  & 21.45 &  Pec/Merger & Binary nucleus ? \\  
% $ 0.779 $ & Pec/Merger   & $37$ m $01$ s $51$ & $11$'$29$"$0$ &
% \tablenotemark{k?} \\
F37192\_1143 & 0.784   &  22.81 &   Sbc & \\
% $ 0.784 $ & Sbc          & $37$ m $19$ s $28$ & $11$'$43$"$5$ & \\
F37083\_1320 & 0.785  &  22.86 & ``Tadpole'' & \\ 
% $ 0.785 $ & ``Tadpole''  & $37$ m $08$ s $33$ & $13$'$20$"$8$ & \\
F37222\_1124 & 0.786 & 22.33 &  Sa & \\
% $ 0.786 $ & Sa           & $37$ m $22$ s $24$ & $11$'$24$"$1$ & \\
F37088\_1214 & 0.788  &  23.90 &  E0/Star & \\
% $ 0.788 $ & E0/Star      & $37$ m $08$ s $83$ & $12$'$14$"$4$ & \\
F37105\_1141 &  0.789  &  21.20 &  Sbcp & No spiral arms \\
% $ 0.789 $ & Sbcp         & $37$ m $10$ s $56$ & $11$'$41$"$0$ &
% \tablenotemark{u}  \\
H36555\_1245 &  0.790   & 23.08 &  Sbcp & Rudimentary spiral structure \\
% $ 0.790 $ & Sbcp         & $36$ m $55$ s $59$ & $12$'$46$"$2$ &
% \tablenotemark{h}  \\
F36299\_1403 & 0.793  &   21.97 &   Merger & Merger of E0t + ? \\
% $ 0.793 $ & Merger       & $36$ m $29$ s $98$ & $14$'$03$"$1$ &
% \tablenotemark{z}  \\
F36270\_1509 & 0.794 &   21.60 &  E2 & \\
% $ 0.794 $ & E2           & $36$ m $27$ s $07$ & $15$'$09$"$4$ & \\
F36194\_1428 &  0.798  &  22.60 &  Pec  &  \\ 
% $ 0.798 $ & Pec          & $36$ m $19$ s $41$ & $14$'$28$"$7$ & \\
F37007\_1107 &   0.801 &  22.94 & Sab: & \\
% $ 0.801 $ & Sab:         & $37$ m $00$ s $73$ & $11$'$07$"$3$ & \\
F36175\_1402 & 0.818 &  21.73 & Pec & \\
% $ 0.818 $ & Pec          & $36$ m $17$ s $60$ & $14$'$02$"$2$ & \\
H36503\_1418 &  0.819  & 23.41 & Sbp &  \\
% $ 0.819 $ & Sbp          & $36$ m $50$ s $34$ & $14$'$18$"$5$ & \\
F37167\_1042 &  0.821 & 21.59 & Ir/Merger \\
% $ 0.821 $ & Ir/Merger    & $37$ m $16$ s $70$ & $10$'$42$"$2$ & \\
F37141\_1044 &   0.821 & 22.32 & E2 & \\
% $ 0.821 $ & E2           & $37$ m $14$ s $12$ & $10$'$44$"$4$ & \\
F37083\_1252 &  0.838  & 22.20 & Sbp & No spiral arms, spectrum may \\
       &   &   &   &  include Sb: companion \\
% $ 0.838 $ & Sbp          & $37$ m $08$ s $31$ & $12$'$52$"$5$ &
% \tablenotemark{u}  \\
F37083\_1514  &  0.839   & 21.62 & Sp & \\
% $ 0.839 $ & Sp           & $37$ m $08$ s $36$ & $15$'$14$"$6$ & \\
F37065\_1512 & 0.840  & 22.94 & E0/Star & \\
% $ 0.840 $ & E0/Star      & $37$ m $06$ s $54$ & $15$'$12$"$5$ & \\
F37064\_1518  &  0.840 & 22.22 & ? & Asymetric \\
% $ 0.840 $ & ?            & $37$ m $06$ s $27$ & $15$'$17$"$7$ &
% \tablenotemark{f}  \\
F37105\_1116 & 0.841 & 23.60 & Sb(p?) & \\
% $ 0.841 $ & Sb(p?)       & $37$ m $10$ s $54$ & $11$'$16$"$3$ & \\
F36447\_1455 &   0.845  & 22.97 &  Pec & \\
% $ 0.842 $ & Pec          & $36$ m $44$ s $80$ & $14$'$55$"$2$ & \\
F36417\_0943 &   0.845 & 22.51 & Sb & \\
% $ 0.845 $ & Sb           & $36$ m $41$ s $74$ & $09$'$43$"$3$ & \\
F36425\_1121 &  0.845  & 23.03 & Sbp & \\
% $ 0.845 $ & Sbp          & $36$ m $42$ s $55$ & $11$'$21$"$9$ & \\
F36343\_1312 &  0.845 &  23.15 &  Pec/Merger? & \\
% $ 0.845 $ & Pec/Merger?  & $36$ m $34$ s $36$ & $13$'$12$"$5$ & \\
F36336\_1319 &   0.845 & 21.78 &  Scp & No spiral arms \\
% $ 1.015 $ & Scp          & $36$ m $33$ s $60$ & $13$'$19$"$8$ &
% \tablenotemark{u?} \\
%  typo in redshift
% F36420\_1321  &  0.846 & 23.95 & E1 & \\
% $ 0.842 $ & E1           & $36$ m $42$ s $04$ & $13$'$21$"$0$ & \\
%  appears to be a duplicate entry/ found 4/25/2001
F36420\_1321 &  0.846 &  23.95 & Sap &  Asymmetric envelope \\
% $ 0.846 $ & Sap          & $36$ m $42$ s $04$ & $13$'$21$"$2$ &
% \tablenotemark{d}  \\
F36341\_1305 &  0.847 & 24.24 & Sb: & \\
% $ 0.847 $ & Sb:          & $36$ m $34$ s $17$ & $13$'$05$"$8$ & \\
F36398\_1249  &  0.848  & 21.53 & Sa: & Nucleus too small for class \\
% $ 0.848 $ & Sa:          & $36$ m $38$ s $92$ & $12$'$50$"$0$ &
% \tablenotemark{x}  \\
F36176\_1408  &  0.848 & 22.55 & Sp & \\
% $ 0.848 $ & Sp           & $36$ m $17$ s $65$ & $14$'$08$"$1$ & \\
H36431\_1242 &   0.849 & 22.34 & E2 & \\
% $ 0.849 $ & E2           & $36$ m $43$ s $16$ & $12$'$42$"$2$ & \\
F36570\_1511 & 0.849 & 23.50 & E1: & \\
% $ 0.849 $ & E1:          & $36$ m $57$ s $07$ & $15$'$11$"$2$ & \\
F36541\_1514 &  0.849  & 22.84 & Pec & \\
% $ 0.849 $ & Pec          & $36$ m $54$ s $13$ & $15$'$14$"$9$ & \\
H36504\_1315 &  0.851 & 23.41 & Sbp & \\
% $ 0.851 $ & Sbp          & $36$ m $50$ s $48$ & $13$'$16$"$1$ & \\
H36540\_1354 &   0.851 &  22.72 & Sc? & \\
% $ 0.851 $ & Sc?          & $36$ m $54$ s $07$ & $13$'$54$"$2$ & \\
F36462\_1527 &  0.851  & 22.11 & Ir & \\
% $ 0.851 $ & Ir           & $36$ m $46$ s $22$ & $15$'$27$"$3$ & \\
F36539\_1606 &  0.851  & 22.84 & Pec & Non-elliptical isophotes \\
% $ 0.851 $ & Pec          & $36$ m $53$ s $90$ & $16$'$06$"$9$ &
% \tablenotemark{p}  \\
F36589\_1208 &  0.853 &  22.32 & Sbp &  Off-center nucleus \\
% $ 0.853 $ & Sbp          & $36$ m $58$ s $96$ & $12$'$08$"$8$ &
% \tablenotemark{t}  \\
F37114\_1054 &   0.855 & 22.44 & Sb & \\
% $ 0.855 $ & Sb           & $37$ m $11$ s $46$ & $10$'$55$"$3$ & \\
F37089\_1202 &  0.855  & 22.90 & Scp & Knots but no arms \\
% $ 0.855 $ & Scp          & $37$ m $08$ s $95$ & $12$'$02$"$2$ &
% \tablenotemark{v}  \\
F37129\_1028 &  0.858  & 22.62 & Pec: &  Edge-on \\
% $ 0.858 $ & Pec:         & $37$ m $12$ s $96$ & $10$'$28$"$6$ &
% \tablenotemark{e}  \\
F37096\_1055 &  0.858 &  23.16 & Pec \\
% $ 0.858 $ & Pec          & $37$ m $09$ s $61$ & $10$'$55$"$2$ & \\
F37041\_1239 &  0.861 &  23.16 & Pec & \\
% $ 0.861 $ & Pec          & $37$ m $04$ s $20$ & $12$'$39$"$6$ & \\
F36472\_1628 &  0.873 & 21.69 & Sa & \\
% $ 0.873 $ & Sa           & $36$ m $47$ s $28$ & $16$'$28$"$3$ & \\
F36408\_1054 &  0.875 & 22.61 & Merger? & \\
% $ 0.875 $ & Merger?      & $36$ m $40$ s $88$ & $10$'$54$"$7$ & \\
H36441\_1240 &  0.875  &  23.39 & Pec & \\
% $ 0.875 $ & Pec          & $36$ m $44$ s $19$ & $12$'$40$"$3$ & \\
F36287\_1239 &  0.880  & 22.11 & Sa: & \\
% $ 0.880 $ & Sa:          & $36$ m $28$ s $80$ & $12$'$39$"$3$ & \\
H36408\_1205 &  0.882  & 22.94 & E1p & Asymmetric envelope \\
% $ 0.882 $ & E1p          & $36$ m $40$ s $94$ & $12$'$05$"$3$ &
% \tablenotemark{d}  \\
F36482\_1507  &  0.890 & 22.38 & Sa & \\
% $ 0.892 $ & Sa           & $36$ m $48$ s $28$ & $15$'$07$"$4$ & \\
F37029\_1427 &   0.898 & 23.67 & E:1 & \\
% $ 0.898 $ & E:1          & $37$ m $02$ s $92$ & $14$'$27$"$6$ & \\
H36461\_1246 &  0.900  & 22.86 & E1 & \\
% $ 0.900 $ & E1           & $36$ m $46$ s $13$ & $12$'$46$"$5$ & \\
F37058\_1153 &   0.904 & 21.22 & Scp & \\
% $ 0.904 $ & Scp          & $37$ m $05$ s $84$ & $11$'$53$"$8$ & \\
H36386\_1233 &  0.904  & 24.04 & Sab & Edge-on \\
% $ 0.904 $ & Sab          & $36$ m $38$ s $61$ & $12$'$33$"$8$ &
% \tablenotemark{e}  \\
F36469\_0906 &  0.905  & 23.84 & Sbp & Asymetric, image too small \\
             &         &       &     &  to classify with confidence \\
% $ 0.905 $ & Sbp          & $36$ m $46$ s $94$ & $09$'$06$"$6$ &
% \tablenotemark{f} \tablenotemark{w} \\
H36501\_1216  &  0.905  & 23.06 & proto-Sbc? & Edge-on \\
% $ 0.905 $ & proto-Sbc?   & $36$ m $50$ s $15$ & $12$'$16$"$9$ &
% \tablenotemark{e}  \\
F37176\_1113 &   0.906  & 22.01 & E3/Sa & \\
% $ 0.906 $ & E3/Sa        & $37$ m $17$ s $65$ & $11$'$14$"$2$ & \\
F37086\_1128 &  0.907  & 22.23 & Merger & \\
% $ 0.907 $ & Merger       & $37$ m $08$ s $66$ & $11$'$28$"$5$ & \\
F37196\_1256 &  0.909  & 23.31 & Sab & Asymetric \\
% $ 0.909 $ & Sab          & $37$ m $19$ s $60$ & $12$'$56$"$2$ &
% \tablenotemark{f}  \\
F37180\_1248 &  0.912 & 22.89 & Sbc & \\
% $ 0.912 $ & Sbc          & $37$ m $18$ s $05$ & $12$'$48$"$3$ & \\
F36468\_1540 &  0.912  & 22.23 & Sc(p?) & \\
% $ 0.912 $ & Sc(p?)       & $36$ m $46$ s $84$ & $15$'$40$"$6$ & \\
F37003\_1616  &  0.913  & 22.33 & Merger?  & \\
% $ 0.913 $ & Merger?      & $37$ m $00$ s $36$ & $16$'$16$"$9$ & \\
F37001\_1615  & 0.914 & 22.83 & Sab & \\
% $ 0.914 $ & Sab          & $37$ m $00$ s $16$ & $16$'$15$"$1$ & \\
F36518\_1125 &  0.919 & 21.62 & Ir/Merger? \\
% $ 0.919 $ & Ir/Merger?   & $36$ m $51$ s $84$ & $11$'$25$"$4$ & \\
H36566\_1220 &   0.930  & 23.15 & E2p & \\
% $ 0.930 $ & E2p          & $36$ m $56$ s $61$ & $12$'$20$"$1$ & \\
F37133\_1054 &  0.936 &  21.87 & E2p & Asymmetric \\
% $ 0.936 $ & E2p          & $37$ m $13$ s $30$ & $10$'$54$"$3$ &
% \tablenotemark{f}  \\
F36522\_1537 &  0.936  & 22.74 & E0/Star \\
% $ 0.936 $ & E0/Star      & $36$ m $52$ s $30$ & $15$'$37$"$0$ & \\
F36459\_1101 &  0.936 &  22.77 & E0 & \\
% $ 0.936 $ & E0           & $36$ m $45$ s $97$ & $11$'$01$"$2$ & \\
F37078\_1605 & 0.936 &  21.88 & proto-Sc \\
% $ 0.936 $ & proto-Sc     & $37$ m $07$ s $84$ & $16$'$05$"$7$ & \\
F36444\_1052 & 0.937 &  23.50 & ``Tadpole?''& Might also be an asymmetric Sab \\
% $ 0.937 $ & ``Tadpole?'' & $36$ m $44$ s $42$ & $10$'$52$"$8$ &
% \tablenotemark{ab} \\
F37018\_1509  &   0.938   & 22.20 & Sbp & \\
% $ 0.938 $ & Sbp          & $37$ m $01$ s $92$ & $15$'$10$"$3$ & \\
F36532\_1116 &  0.942 &  22.08 & Proto-Sc? \\
% $ 0.942 $ & Proto-Sc?    & $36$ m $53$ s $20$ & $11$'$16$"$9$ & \\
F36529\_1508 &  0.942 &  22.84 &  E0/Star \\
% $ 0.942 $ & E0/Star      & $36$ m $52$ s $99$ & $15$'$08$"$6$ & \\
H36396\_1230 &  0.943  &  24.40 &  E/Sa(p?) \\
% $ 0.943 $ & E/Sa(p?)     & $36$ m $39$ s $60$ & $12$'$30$"$2$ & \\
H36384\_1231 &  0.944  & 22.87 & Amorphous & Edge-on, off center nucleus ? \\
% $ 0.944 $ & Amorphous    & $36$ m $38$ s $43$ & $12$'$31$"$2$ &
% \tablenotemark{e}  \\
F36465\_1049 & 0.945 &  23.70 & Sa? & \\
% $ 0.857 $ & Sa?          & $36$ m $46$ s $56$ & $10$'$49$"$1$ & \\
% redshift changed 4/20/2001, see Adelberger & Steidel (2000)
H36555\_1249 &   0.950  & 23.53 & proto-Sc & \\
% $ 0.950 $ & proto-Sc     & $36$ m $55$ s $59$ & $12$'$49$"$3$ & \\
H36551\_1303  &  0.952   & 24.29 & E3 & \\
% $ 0.952 $ & E3           & $36$ m $55$ s $14$ & $13$'$03$"$7$ & \\
H36576\_1315 &   0.952  & 22.94 & Pec/Merger & \\
% $ 0.952 $ & Pec/Merger   & $36$ m $57$ s $69$ & $13$'$15$"$3$ & \\
H36490\_1221 &  0.953 & 22.59 & Pec & Nucleus + 4 bright knots \\
% $ 0.953 $ & Pec          & $36$ m $49$ s $05$ & $12$'$21$"$1$ &
% \tablenotemark{r}  \\
F36524\_0919 &  0.954  & 22.81 & E3 \\
% $ 0.954 $ & E3           & $36$ m $52$ s $49$ & $09$'$19$"$6$ & \\
F36502\_1127 &  0.954 &  22.88 & Sbcp & \\
% $ 0.954 $ & Sbcp         & $36$ m $50$ s $27$ & $11$'$27$"$4$ & \\
F36520\_1059 & 0.955 & 23.67 & S & \\
% $ 0.955 $ & S            & $36$ m $52$ s $03$ & $10$'$59$"$1$ & \\
H36486\_1328 &  0.958 &  23.14 & Pec/Merger & \\
% $ 0.958 $ & Pec/Merger   & $36$ m $48$ s $62$ & $13$'$28$"$1$ & \\
H36477\_1232 &  0.960 &  23.80 & Sa & \\
% $ 0.960 $ & Sa           & $36$ m $47$ s $79$ & $12$'$32$"$9$ & \\
F36366\_1346  &  0.960  & 20.32 & E0 & \\
% $ 0.962 $ & E0           & $36$ m $36$ s $65$ & $13$'$46$"$6$ & \\
H36492\_1148 &  0.961 & 23.26 & Sa:p & Asymmetric \\
% $ 0.961 $ & Sa:p         & $36$ m $49$ s $24$ & $11$'$48$"$8$ &
% \tablenotemark{f}  \\
H36493\_1155 &   0.961 &  23.36 & E2 & \\
% $ 0.961 $ & E2           & $36$ m $49$ s $34$ & $11$'$55$"$1$ & \\
F36364\_1237 &  0.961  & 22.94 & Sbp & \\
% $ 0.961 $ & Sbp          & $36$ m $36$ s $40$ & $12$'$37$"$3$ & \\
F36486\_1141 &  0.962  & 22.21 & Sbt & \\
% $ 0.962 $ & Sbt          & $36$ m $48$ s $61$ & $11$'$41$"$1$ & \\
H36483\_1214 &   0.962  &  23.87 & S(B)bct & \\
% $ 0.962 $ & S(B)bct      & $36$ m $48$ s $33$ & $12$'$14$"$3$ & \\
H36463\_1404 &  0.962  & 21.69 & E0 & Embedded in large shell \\
% $ 0.962 $ & E0           & $36$ m $46$ s $34$ & $14$'$04$"$6$ &
% \tablenotemark{b}  \\
F37224\_1216 &   0.963 & 22.23 & Sb(t?) & Part of multiple interacting system \\
% $ 0.963 $ & Sb(t?)       & $37$ m $22$ s $48$ & $12$'$16$"$0$ &
% \tablenotemark{y}  \\
H36554\_1310  &  0.968  & 22.86 & E1 & \\
% $ 0.968 $ & E1           & $36$ m $55$ s $39$ & $13$'$11$"$0$ & \\
F37058\_1423 &   0.970   & 22.48 & Sb: & \\
% $ 0.970 $ & Sb:          & $37$ m $05$ s $83$ & $14$'$23$"$4$ & \\
F37055\_1129 &  1.001 &  22.37 & Sap & \\
% $ 1.001 $ & Sap          & $37$ m $05$ s $57$ & $11$'$29$"$2$ & \\
H36432\_1148 &  1.010  & 23.10 & Sbp & Smooth disk with little star formation ? \\
% $ 1.010 $ & Sbp          & $36$ m $43$ s $21$ & $11$'$48$"$1$ &
% \tablenotemark{m?} \\
H36408\_1203  &  1.010  & 23.49 & Ir & Edge-on \\
% $ 1.010 $ & Ir           & $36$ m $40$ s $85$ & $12$'$03$"$1$ &
% \tablenotemark{e}  \\
H36461\_1142 &  1.013  & 21.52 & proto-Sc & \\
% $ 1.013 $ & proto-Sc     & $36$ m $46$ s $17$ & $11$'$42$"$2$ & \\
F37000\_1605  &  1.013  & 22.46 & Sc: & On edge of image \\
% $ 1.013 $ & Sc:          & $37$ m $00$ s $04$ & $16$'$05$"$7$ &
% \tablenotemark{a}  \\
F37154\_1212 &  1.014  & 23.25 & Ep & \\
% $ 1.014 $ & Ep           & $37$ m $15$ s $49$ & $12$'$12$"$2$ & \\
H36400\_1207  &  1.015 & 22.75 & E0 & \\
% $ 1.015 $ & E0           & $36$ m $40$ s $02$ & $12$'$07$"$3$ & \\
F36411\_1314 &  1.017  &  23.08 & Sap & Asymmetric \\
% $ 1.017 $ & Sap          & $36$ m $41$ s $13$ & $13$'$14$"$3$ &
% \tablenotemark{f}  \\
F36497\_1106 & 1.018 & 23.37 & Pec & \\
% $ 1.018 $ & Pec          & $36$ m $49$ s $75$ & $11$'$06$"$7$ & \\
H36444\_1142 &  1.020  & 24.30 & Pec/Merger \\
% $ 1.020 $ & Pec/Merger   & $36$ m $44$ s $49$ & $11$'$42$"$3$ & \\
F37159\_1213 &  1.020 &  23.27 & Sbcp & \\
% $ 1.020 $ & Sbcp         & $37$ m $15$ s $91$ & $12$'$13$"$3$ & \\
F36583\_1214 &  1.020  & 23.79 & E2 \\
% $ 1.020 $ & E2           & $36$ m $58$ s $35$ & $12$'$14$"$1$ & \\
F36595\_1153 & 1.021 &   22.54 & Sap(t?) & Asymmetric \\
% $ 1.021 $ & Sap(t?)      & $36$ m $59$ s $54$ & $11$'$53$"$9$ &
% \tablenotemark{f}  \\
H36443\_1133 & 1.050 &  21.96 & E1 & \\
% $ 1.050 $ & E1           & $36$ m $44$ s $36$ & $11$'$33$"$2$ & \\
F37046\_1415 &  1.050 & 23.97 & Sab: & \\
% $ 1.052 $ & Sab:         & $37$ m $04$ s $61$ & $14$'$16$"$0$ & \\
F36296\_1420 &  1.055   & 24.00 & E0/Star & \\
% $ 1.055 $ & E0/Star      & $36$ m $29$ s $68$ & $14$'$20$"$6$ & \\
H36467\_1144 &  1.060  & 24.23 & proto-Sc & \\
% $ 1.060 $ & proto-Sc     & $36$ m $46$ s $80$ & $11$'$44$"$9$ & \\
F37026\_1216 &  1.073  & 24.04 & Sabp & Asymmetric \\
% $ 1.073 $ & Sabp         & $37$ m $02$ s $60$ & $12$'$16$"$6$ &
% \tablenotemark{f}  \\
F37143\_1221 &   1.084   & 24.12 & E:2 & Image too small to classify with confidence \\
% $ 1.084 $ & E:2          & $37$ m $14$ s $38$ & $12$'$21$"$2$ &
% \tablenotemark{w}  \\
H36519\_1332 &   1.087  &  23.59 & Sap & \\
% $ 1.087 $ & Sap          & $36$ m $51$ s $96$ & $13$'$32$"$1$ & \\
\enddata
\tablenotetext{a}{Names are Habcde\_fghi for objects in the HDF, where the object's
J2000 coordinates are 12 ab cd.e +62 fg hi.  The initial letter is
``F'' for objects in the flanking fields.}
\tablenotetext{b}{Redshifts are from Cohen \etal\ (2000) or Cohen (2001).}
\tablenotetext{c}{$R$ magnitudes are from Hogg \etal\ (2000).}
\end{deluxetable}

\begin{deluxetable}{lllcllllllll}
\tablewidth{0pt}
\tablecaption{ \sc{Frequency Distribution of Galaxy Types} }
\tablehead{ \colhead{$ z $} & \colhead{$ n(z) $} &
\colhead{E\tablenotemark{a}} & \colhead{Sa + Sab} & \colhead{Sb +
Sbc\tablenotemark{b} } & \colhead{Sc + Sc/Ir\tablenotemark{b}} & \colhead{Ir}
& \colhead{Pec} & \colhead{P/Mer} & \colhead{Mer} & \colhead{?} }
\startdata
$ 0.20-0.29 $ & $  \phn \phn 4 $ & $  \phn 0 $ & $  \phn 1 $ & $  \phn 1 $
& $  \phn 0 $      & $  \phn 2   $ & $  0 $ & $ 0 $ & $ 0   $ & $ 0 $ \\
$ 0.30-0.39 $ & $  \phn \phn 7 $ & $  \phn 0 $ & $  \phn 1 $ & $  \phn 2.5 $
& $  \phn 1 $      & $  \phn 0.5   $ & $  1 $ & $ 0 $ & $ 1   $ & $ 0 $ \\
$ 0.40-0.49 $ & $  \phn 25 $     & $  \phn 4 $ & $  \phn 4 $ & $  \phn 5 $
& $  \phn 4+(1)  $ & $  \phn 0   $ & $  2 $ & $ 1 $ & $ 0   $ & $ 0 $ \\
$ 0.50-0.59 $ & $  \phn 24 $     & $  \phn 5 $ & $  \phn 2 $ & $  \phn 5 $
& $  \phn 3      $ & $  \phn 2.5 $ & $  0 $ & $ 1 $ & $ 1   $ & $ 1 $ \\
$ 0.60-0.69 $ & $  \phn 37 $     & $ 14 $      & $  \phn 7 $ & $  \phn 7 $
& $  \phn 1+(1)  $ & $  \phn 1.5 $ & $  1 $ & $ 0 $ & $ 0   $ & $ 0 $ \\
$ 0.70-0.79 $ & $  \phn 32 $     & $  \phn 4 $ & $  \phn 7 $ & $  \phn 8 $
& $  \phn 2+(1)  $ & $  \phn 1   $ & $  1 $ & $ 1 $ & $ 2.5 $ & $ 2 $ \\
$ 0.80-0.89 $ & $  \phn 41 $     & $  \phn 7 $ & $  \phn 7 $ & $ 9+(1) $
& $  \phn 3      $ & $  \phn 1.5 $ & $  9 $ & $ 1 $ & $ 2   $ & $ 1 $ \\
$ 0.90-0.99 $ & $  \phn 43 $     & $ 14 $      & $  \phn 5 $ & $ 11 $
& $  \phn 2+(2)  $ & $  \phn 0.5 $ & $  1 $ & $ 2 $ & $ 2.5 $ & $ 0 $ \\
$ 1.00-1.09 $ & $  \phn 20 $     & $  \phn 6 $ & $  \phn 6 $ & $  \phn 0 $
& $  \phn 1+(1)  $ & $  \phn 1   $ & $  1 $ & $ 1 $ & $ 0   $ & $ 0 $ \\
\hline
Total         & $ 233 $ & $ 55 $ & $ 40 $ & $ 48.5+(1) $ & $ 17+(6)  $ & $ 10.5
$ & $ 16 $ & $ 7 $ & $ 9   $ & $ 4 $ \\
\enddata
\tablenotetext{a}{Includes E, E0/Star, E/Sa, S0 and S0/Sa}
\tablenotetext{b}{Numbers in parentheses refer to objects that were
classified as proto-Sb or proto-Sc}
\end{deluxetable}

\clearpage

\begin{figure}
\plotone{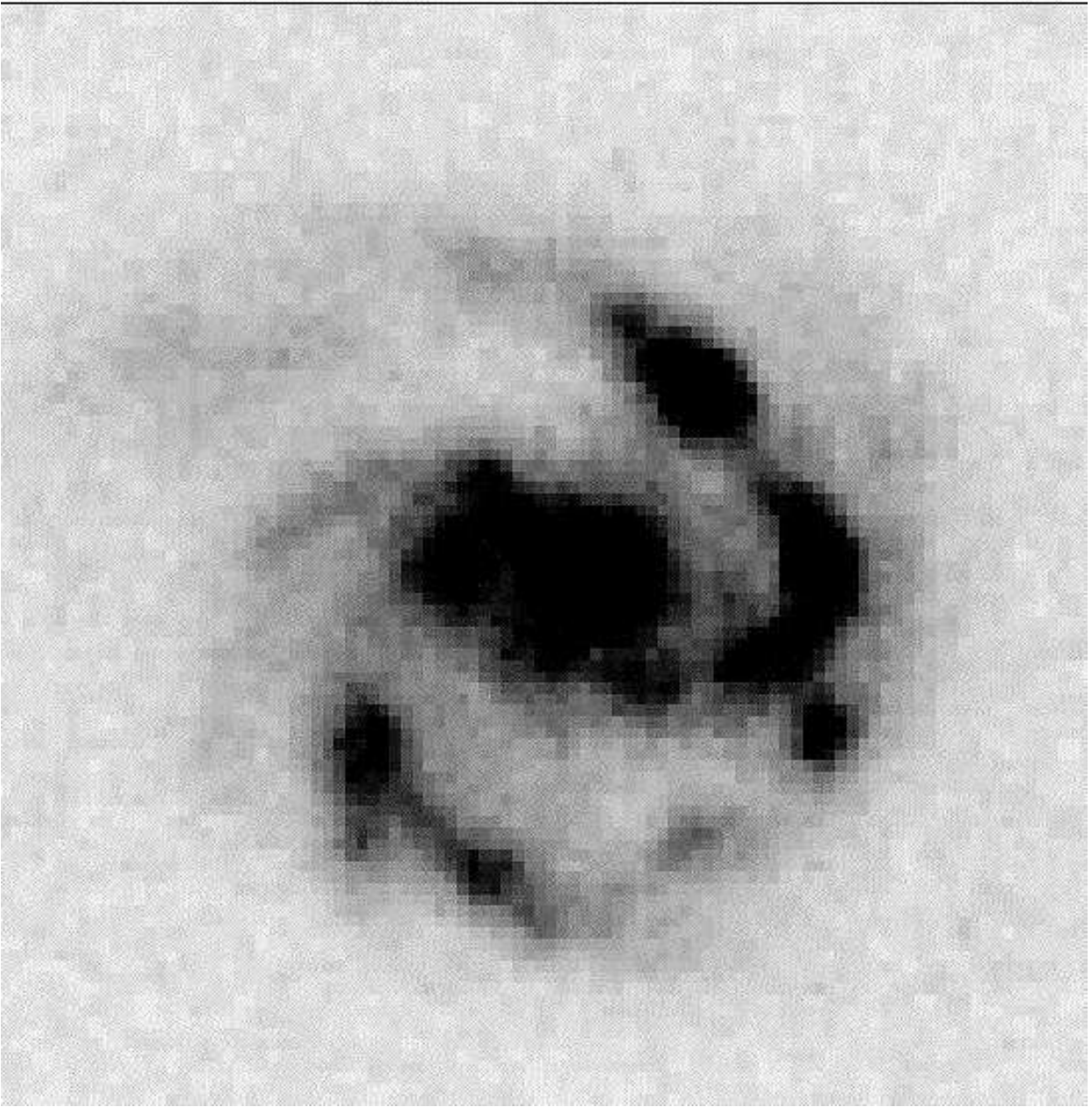}
\caption{Example of what is interpreted as a proto-Sc 
         galaxy. Note the rather chaotic spiral structure 
         and very patchy arm morphology of H36461\_1142 at 
         $z=1.01$.  (Original figures have slightly more contrast.)}
\end{figure}

\begin{figure}
\plotone{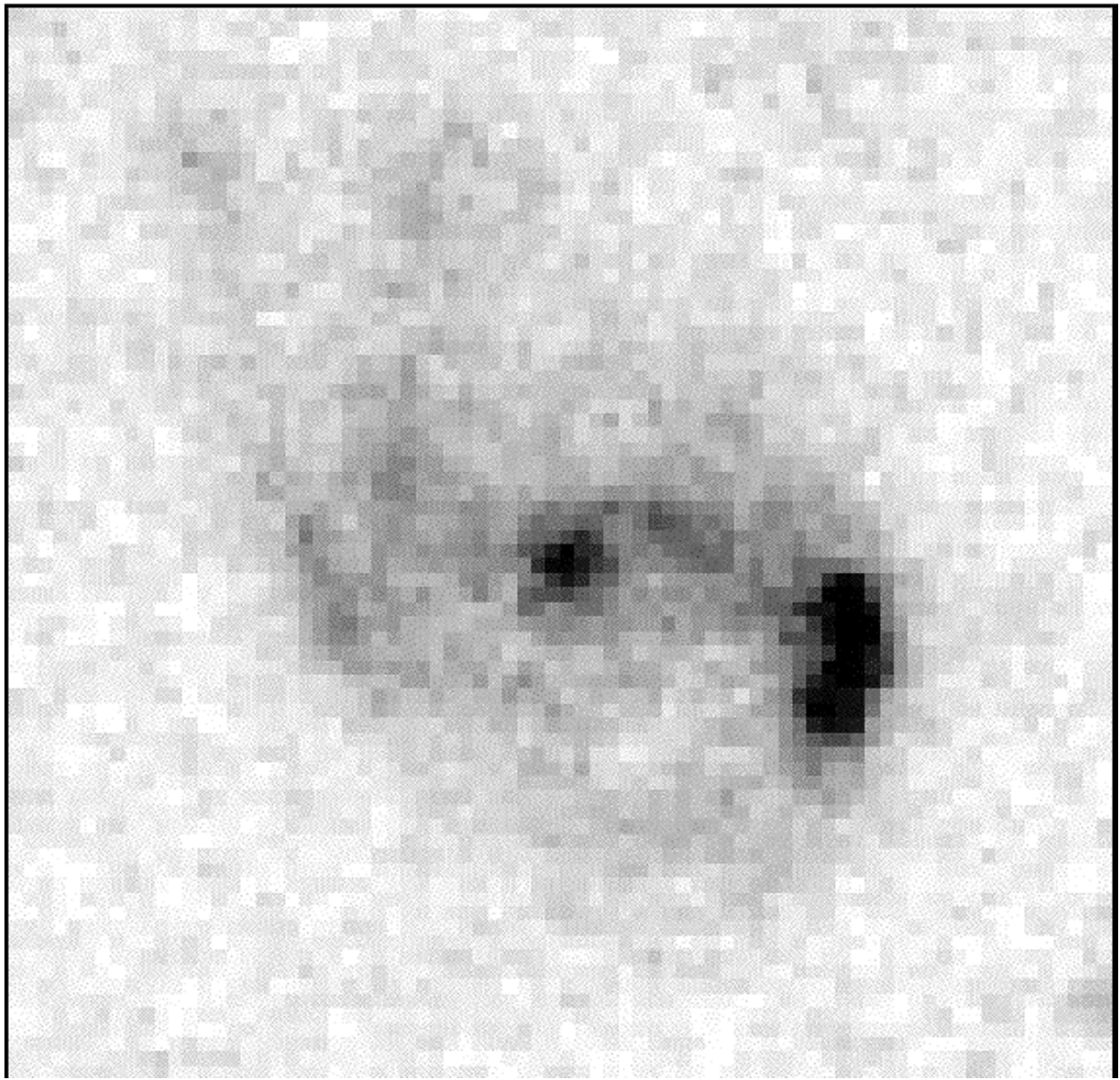}
\caption{This spiral(?) (H36467\_1144)
might be interpreted as either a
         proto-Sc with a very patchy arm structure, or
         as the product of a recent merger. Kinematical
         studies of this object at $z = 1.06$  will
         be required to distinguish between these two
         alternatives.  (Original figures have slightly more contrast.)}
\end{figure}

\begin{figure}
\plotone{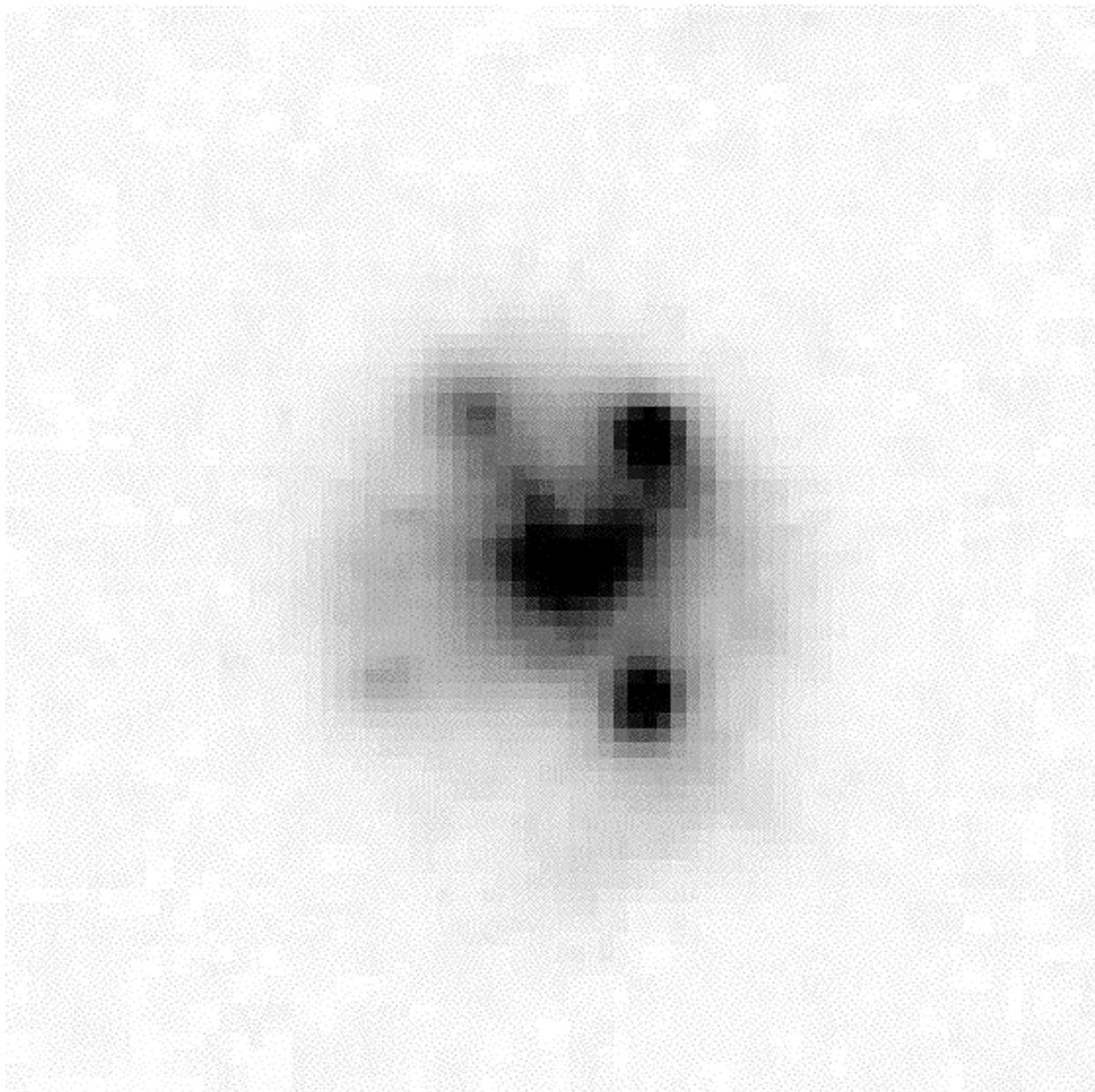}
\caption{This peculiar disk galaxy (H36490\_1221)
at $z=0.953$ consists of a disk in which five bright knots,
         one of which might be the nucleus, are embedded.
         No spiral structure appears to have developed
         yet in this object.  (Original figures have slightly more contrast.)}
\end{figure}


\clearpage

\begin{thebibliography}{}

\bibitem [Abraham (1999)]{abraham1999}Abraham, R. G., 1999, \apss, 269--70,
323

\bibitem [Abraham \etal\ (1999)]{abrahametal1999}Abraham, R. G., Merrifield,
M. R., Ellis, R. S., Tanvir, N. R., \& Brinchmann, J., 1999, \mnras, 308,
569

\bibitem [Abraham \& van den Bergh (in preparation)]{abrahamvdbprep}Abraham,
R. G. \& van den Bergh, S., 2001, in preparation

\bibitem [Binchmann \etal\ (1998)]{binchmannetal1998}Binchmann, J. \etal\,
1998, \apj, 499, 112

\bibitem [Bothun (2000)]{bothun2000}Bothun, G. D., 2000, \skytel, May, 37

\bibitem [Carollo \& Lilly (2001)]{carollolilly2001}Carollo, C. M. \& Lilly,
S. J., 2001, \apj, 548, L153

\bibitem [Cohen (2001)]{cohen2001}Cohen, J. G., 2001, \aj, in press
(see Astro-ph/0101251)

\bibitem [Cohen (2000)]{cohen2000}Cohen, J. G., 2000, to appear in
{\it Deep Fields}, ed. S.Cristiani, A.Renzini \& R.E.Williams,
ESO Astrophysics Symposia, Springer-Verlag, Heidelberg (see Astro-ph/0012004)

\bibitem [Cohen \etal\ (2000)]{cohenetal2000}Cohen, J. G., Hogg, D. W.,
Blandford, R. D., Cowie, L. L.,
Hu, E., Songaila, A., Shopbell, P. \& Richberg, K., 2000, \apj, 538, 29

\bibitem [Drozdovsky \etal\ (2001)]{drozdovskyetal2001}Drozdovsky, I. O.,
Schulte-Ladbeck, R. E., Hopp, U., Crone, M. M. \& Greggio, L., 2001, \apj,
in press (see Astro-ph/0102452)

\bibitem [Eskridge \etal\ (2000)]{eskridgeetal2000}Eskridge, P. B. \etal\,
2000, \aj, 119, 536

\bibitem [Guzm\'an \etal\ (1997)]{guzmanetal1997}Guzm\'an, R., Gallego, J.,
Koo, D. C., Phillips, A. C., Lowenthal, J. D., Faber, S. M., Illingworth, G.
D., \& Vogt, N. P., 1997, \apj, 489, 559

\bibitem [Hubble (1936)]{hubble1936}Hubble, E., 1936, {\it{The Realm of the
Nebulae}}, (New
Haven: Yale University Press), p.45

\bibitem [Lilly \etal\ (1995)]{lillyetal1995}Lilly, S. J., Tresse, L.,
Hammer, F., Crampton, D., \& Le F\'evre, O., 1995, ApJ, 455, 108

\bibitem [Noguchi (1998)]{noguchi1998}Noguchi, M., 1998, \nat, 392, 253

\bibitem [Phillips \etal\ (1997)]{phillipsetal1997}Phillips, A. C.,
Guzm\'an, R., Gallego, J., Doo, D. C., Lowenthal, J. D., Vogt, N. P., Faber,
S. M., \& Illingworth, S.D., 1997, \apj, 489, 543

\bibitem [Sandage \& Tammann (1981)]{sandagetammann1981}Sandage, A. \&
Tammann, G. A., 1981,
{\it{A Revised Shapley-Ames Catalog of Bright Galaxies}}, (Washington:
Carnegie Institution)

\bibitem [van den Bergh (1960a)]{vdb1960a}van den Bergh, S., 1960a, \apj,
131, 215

\bibitem [van den Bergh (1960b)]{vdb1960b}van den Bergh, S., 1960b, \apj,
131, 558

\bibitem [van den Bergh (1960c)]{vdb1960c}van den Bergh, S., 1960c, Pub.
David Dunlap Obs., 2, 159

\bibitem [van den Bergh (1989)]{vdb1989}van den Bergh, S., 1989, \aj, 97,
1556

\bibitem [van den Bergh (1998)]{vdb1998}van den Bergh, S., 1998, {\it{Galaxy
Morphology and 
Classification}}, (Cambridge: Cambridge Univ. Press)

\bibitem [van den Bergh \etal\ (1996)]{vdbetal1996}van den Bergh, S.,
Abraham, R. G., Ellis, R. S., Tanvir, N.
R., Santiago, B. X. \& Glazebrook, K. G., 1996, \aj, 112, 359

\bibitem [van den Bergh \etal\ (2000)]{vdbetal2000}van den Bergh, S., Cohen,
J. G., Hogg, D. W., \& Blandford, R., 2000, \aj, 120, 2190

\bibitem [Williams \etal\ (1996)]{williamsetal1996}Williams, R. E. \etal\,
1996, \aj, 112, 1335

\end{thebibliography}
\end{document}